\documentclass[
hf,
]{ceurart}

\usepackage{listings}
\begin{document}

\copyrightyear{2024}
\copyrightclause{Copyright for this paper by its authors.
  Use permitted under Creative Commons License Attribution 4.0
  International (CC BY 4.0).}

\conference{%
}

\title{A Survey of Web Content Control for Generative AI}


\author[1]{Michael Dinzinger}[%
orcid=0009-0003-1747-5643,
email=michael.dinzinger@uni-passau.de,
]
\cormark[1]
\address[1]{University of Passau, Innstraße 41, 94032 Passau, Germany}

\author[1]{Florian Heß}[%
orcid=0009-0008-8284-6988,
email=florian.hess@uni-passau.de,
]

\author[1]{Michael Granitzer}[%
orcid=0000-0003-3566-5507,
email=michael.granitzer@uni-passau.de,
]

\cortext[1]{Corresponding author.}

\begin{abstract}
The groundbreaking advancements around generative AI have recently caused a wave of concern culminating in a row of lawsuits, including high-profile actions against Stability AI and OpenAI. This situation of legal uncertainty has sparked a broad discussion on the rights of content creators and publishers to protect their intellectual property on the web. European as well as US law already provides rough guidelines, setting a direction for technical solutions to regulate web data use. In this course, researchers and practitioners have worked on numerous web standards and opt-out formats that empower publishers to keep their data out of the development of generative AI models. The emerging AI/ML opt-out protocols are valuable in regards to data sovereignty, but again, it creates an adverse situation for a site owners who are overwhelmed by the multitude of recent ad hoc standards to consider. In our work, we want to survey the different proposals, ideas and initiatives, and provide a comprehensive legal and technical background in the context of the current discussion on web publishers control.
\end{abstract}


\maketitle

\thispagestyle{empty}

\section{Introduction}

The recent advancements in text- and image-generating Large Language Models (LLMs), such as ChatGPT, LLaMA, and Stable Diffusion, have brought Artificial Intelligence (AI) into the forefront of our daily lives~\cite{Bubeck2023, Touvron2023, Zhao2023}. This technological breakthrough has sparked widespread excitement in regards to its potential for increased productivity and societal progress. However, it has also raised significant concerns due to its potentially disruptive effects~\cite{Dwivedi2023}. One of the primary issues is that LLMs are trained on vast amounts of data from the web, often collected without the explicit permission of the authors. Furthermore, these models cannot trace back and cite the original sources of the human-generated content they were trained on. This poses a particular problem for publishers and content creators who want to preserve the authenticity and economic value of their original content. As a result, they are increasingly seeking ways to keep their web content out of the training datasets for these models, in an effort to safeguard their intellectual property on the web. The current mechanisms available to enforce such restrictions are however inadequate, leading to a significant gap in web publishers' ability to control how their data is used downstream.

Regulations for online data providers and consumers are a key measure for surmounting this shortcoming, and the existing legal framework already provides a rough direction. It is now up to the web community to establish simple and practical solutions that meet these requirements. New technical standards may fill this gap and empower rightsholders with finer control over how their data shall be used by well-intentioned practitioners. In this context, a \textit{web standard} refers to a technical specification (protocol specification, RfC, etc.) endorsed by a recognized web standards organization such as W3C, IETF, etc. In contrast, \textit{pseudo standards} are widely used specifications not formally recognized by such organizations, while \textit{ad hoc standards} are emerging ideas or practices concerning web technologies that are not yet widely adopted or officially standardized.

There has been a notable effort among researchers and practitioners to bridge the communication gap between data providers and users, leading to the development of various ad hoc standards. These initiatives range from small software solutions to large-scale community projects. Our research aims to contribute to these efforts by evaluating recent approaches and examining the legal and technical aspects of this complex issue. Our work is focused on three main areas:
\vspace{-0.05cm}
\begin{itemize}
\item The legal background around intellectual property and data protection, including EU's 2019 DSM Directive, which offers a regulatory framework for opting out of Text \& Data Mining (Section 2).
\item A comprehensive review of past and present standards for controlling web data usage (Section 3).
\item An in-depth evaluation of recent ad hoc standards and an analysis of their practical application (Section 4).
\end{itemize}

\section{Legal background}

This section details the legal background any standard on web data protection has to be embedded in. The protection itself arises mainly from intellectual property and data protection law. Unfortunately, as a general rule, each country's legislation applies exclusively within its own borders so that this background varies from one country to another. Harmonization has only been archieved up to a limited extent. We therefore focus on EU and US law due to their major global importance.

\subsection{Intellectual property}

\subsubsection{Scope}
Web usage standards are mostly tailored towards intellectual property. 
This comprises copyright and so-called related rights. For copyright protection, EU und US law provide essentially the same criteria. EU law has no codified definition, but the ECJ (European Court of Justice) derives a two-step test from the overall framework~\cite{Infopaq2009}. A work must be, on the hand, sufficiently original (an own intellectual creation representing the person’s personality) and, on the other hand, be an expression of such originality (identifiable with sufficient precision and objectivity)~\cite{Levola2009}. The standard for quality and quantity required for a work to be considered copyright protected is however low, as even as few as eleven words can suffice~\cite{Infopaq2009}. Most content will thus be protected by copyright.

This protection generally starts at the moment of creation and the registration of the corresponding work is not necessary. In the US, registration with the Register of Copyrights is an option but not mandatory for basic protection.
The initial rightsholder is the person who physically brought the work into existence. In copyright terms, this person is referred to as the author, irrelevant of whether they created art, books, music, videos, or any other work. In the area of social media, the initial rightsholders are thus content creators, not host platforms. This applies even if an author has only realized someone else's concept. These rightsholders are conferred certain exclusive rights, inter alia, the right to reproduction, i.e., creating a copy by any means and in any form, no matter how small or limited in time. When someone wishes to reproduce a given work, he needs either the consent of the rightsholder or a copyright exception to apply.

Regardless of the purpose, during web crawling and scraping, the HTML file is downloaded, the full text extracted and analysed. In each case, a local copy and thus a reproduction is created. Justification for the reproductions in the crawling and scraping process is therefore always required. The applicable law depends on the location of the server that hosts the content. Further steps depend on the concrete purpose that may include more reproductions or other implications with copyright. One such case is the training of generative AI models, in which parts of the training data are potentially imitated or reproduced by the model output.

Apart from copyright, intellectual property also includes related rights. These are either based on the creation of new content (e.g. performers) or an economic investment (e.g. databases or press publications). While EU law provides a range of rights, traditionally US law tends to be less expansive.

\subsubsection{European law}
EU copyright law provides various statutory exceptions that used to cover some type of web crawling and scraping but not its entirety~\cite{Elra2018, Truyens2014}.
The remaining cases were decided by national law. For instance, crawling for conventional web indexing was considered legitimate because it is beneficial to all parties involved~\cite{Truyens2014}. The German Supreme Court interpreted this as an \textit{implied consent}, e.g. in \cite{VorschaubilderI}. Websites endorse to a large extent the crawling of Google, Microsoft and alike because they indirectly profit through user traffic that results from displaying the site on search result pages. If an author does not object (e.g. via the Robots Exclusion Protocol), it must be assumed that he wishes to participate in this standard.

This lack for a specific rule was adressed in 2019 when the EU enacted the Directive on Copyright in the Digital Single Market (DSMD). Its Art. 2 (2) DSMD defines Text \& Data Mining (TDM) as ``any automated analytical technique aimed at analysing text and data in digital form in order to generate information which includes but is not limited to patterns, trends and correlations''. This encompasses all types of TDM, including web crawling and scraping, but also further techniques as training an ML model.

Art. 4 DSMD lays down the general rule that there shall be an exception or limitation for reproductions and extractions of lawfully accessible works and other subject matter (see Table~\ref{tab:comparison_dsmd}). Reproductions are thus permissible if carried with lawful access (which does not assume a lawful upload beforehand).
The reproductions may be retained for as long as it is necessary for the purposes of Text \& Data Mining. Rightsholders can object if they expressly reserve the use of the works in an appropriate manner (``opt-out''), e.g. through machine-readable metadata on web pages in case the content has been made publicly available online.

Additionally, Art. 3 DSMD provides specific rules for TDM for the purposes of scientific research. These are more permissive since there is no possibility to object or waive by contract. It applies however just to research organisations and cultural heritage institutions. Research organisations are any entity, the primary goal of which is to conduct scientific research on a not-for-profit basis, by reinvesting all the profits in its scientific research or pursuant to a public interest mission. Cultural heritage institution means a publicly accessible library or museum, an archive or a film or audio heritage institution.

\begin{table}
\centering
\begin{tabular}{ccccc}
     & Purpose & Objections? & Waive by contract?\\
    Art. 4 & Irrelevant & Yes & Yes\\
    Art. 3 & Scientific research & No & No\\
\end{tabular}
\caption{Comparison Art. 3 and 4 DSMD}
\label{tab:comparison_dsmd}
\end{table}

Under the general rule, any web crawling and scraping is thus permissible by default. Authors can however choose to opt out. The problem is that there is not yet a web standard to opt out from certain uses only. The REP, for instance, is a suitable, machine-readable objection, but it merely allows for a complete reservation without distinguishing between different purposes. This may be due to the fact that the domain of crawling was decisively shaped by Search Engines. The protocol has been formed as open standard under the collaboration of Google and consequently evolved towards the requirements of web search and Search Engine Optimization (SEO). With the increasing prevalence of AI technology, rightsholders may still not wish to object to TDM as a whole but only to the training of generative AI. From an EU perspective, web control for authors depends thus on the establishment of more fine-grained standards.

\subsubsection{US law}
Under US law, there are different justifications related to Text \& Data Mining, but the so-called fair use principle in § 107 Title 17 of the US Code is considered to be most suitable. In determining whether the use made of a work is fair, four factors should be weighed:
\begin{itemize}
\item[(1)] the purpose and character of the use,
\item[(2)] the nature of the copyrighted work,
\item[(3)] the effect of the use upon the potential market for or value of the copyrighted work, and
\item[(4)] the amount and substantiality of the portion used in relation to the copyrighted work as a whole.
\end{itemize}

Unlike in EU law, there is no secure mechanism for an opt-out, but courts will do an overall assessment. TDM for Web indexing~\cite{Field2006} or preservation purposes~\cite{AuthorsGuild2014} were considered permissible. Concerning TDM for AI training, the discussion is ongoing and there are still cases to come~\cite{Samuelson2023}. Both OpenAI and Stability AI are currently involved in court proceedings. Rightsholders also dispute the first factor (especially whether the use is transformative, i.e. whether something new is added), but the crux of the matter will most probably lie within the (intertwined) third factor. Reproductions for web indexing or research purposes were decided to have positive impacts on the market. For AI, the dynamics are likely to differ but the debate already starts with question what even the relevant market is: the market for the end product or, beforehand, the market for training data. The latter can at least not be extended too far as theoretically every work can be used as training data.

So while EU law is open to new standards for protection, US law has no statutory need. The situation is still more open.

\subsection{Data protection}
In addition to copyright law, content may be protected by data protection law.

In EU law, any processing of personal data needs a legal basis listed in Art. 6 GDPR. Processing means any operation performed on personal data. It is thus needed a basis for downloading, extracting information, etc., i.e. every single step in web crawling and scraping, but also later, e.g. when training an AI.
This applies not only to processors established within the EU (establishment rule), but also to any processing activities where the data subject is within the EU and the processing is aimed at individuals in the EU (marketplace rule).

Consent could serve as a legal basis but will, most commonly, not be provided. It would have to be given unambiguously, which cannot be assumed by the mere upload. 
Legitimacy therefore depends on a balancing of interests as an alternative legal basis. The interests of the controller or other third parties (including the general public) must be weighed against those of (a person in the place of) the data subject, i.e. the person identified by the information contained in the content (which is not necessarily the author in a copyright sense).
If the data subject has further indiviudual reasons, it can object at any time. It then requires a reevaluation of the particular case.
Unlike in intellectual property, there is thus no direct legal connection for a protocol. The resonating valuation, however, is to be considered in the weighing of interests.
For conventional web indexing there are already differentiated lines of case law (for delisting upon takedown request), but for the training of AI the situation is once again more open.

In the US, there is no comprehensive federal data protection regulation. On a federal level, only certain categories or industries are regulated (e.g. by the Health Insurance Portability and Accountability Act). On a state level, some have enacted further regulations (e.g. the California Consumer Privacy Act). Legality depends thus on the type of data or individual state laws.

\section{Technical background}

The following section introduces the most important practices, ideas and initiatives that are relevant to web data content control.

\subsection{Robots Exclusion Protocol}
In the context of web crawling, one protocol has prevailed as dominating mechanism for content control. The \textit{Robots Exclusion Protocol (REP)} is a common standard among websites and regulates the access of autonomous bots. The protocol encourages webmasters to state access rules for any non-human visitors in a robots.txt file that is placed at the root directory of the web server. These files mainly consist of \texttt{allow} and \texttt{disallow} instructions referring to URL paths, which are structured in groups and assigned to user agents. REP was initially introduced in 1994 by Martijn Koster at Nexor and finally codified as an IETF RFC specification in 2022~\cite{Koster2022}. As a common tool for both webmasters and Search Engine operators, REP restrains disproportinate server traffic and improves the efficiency of crawlers.

The HTML meta tag \texttt{robots} and the HTTP response header \texttt{X-Robots-Tag} declare instructions addressing autonomous web agents, similar to robots.txt. However, the tags are not formally included in the REP standard and furthermore apply to a lower level of structural granularity. Whereas robots.txt relates to the entire site, the embedded robots tags are specific to a single delivered HTML document. Their instructions guide Search Engines how to crawl, index and display information from this particular web page, e.g., nofollow, noindex and max-snippet.

Apparently, the protocol does not specify any enforcement mechanisms. This shortcoming of enforcement given the limited traceability of web agents underlines the central role of trust in the internet. Nonetheless, all major Search Engine operators generally respect the Robots Exclusion standards and previous studies have shown that the protocol is widely adopted among websites~\cite{Kolay2008, Sun2010}.

\subsection{Relevant protocols and initiatives}
Due to its central role, there have been numerous efforts to extend REP and refine the communication between web agents and servers. In 2007, organizatons from the publishing industry proposed the \textit{Automated Content Access Protocol (ACAP)}.\footnote{\url{https://web.archive.org/web/20211011020458/http://the-acap.org/}} The protocol provides Search Engines with detailed information on thumbnail and text snippets to be used for web indexing. The initial version ACAP 1.0 has never been extensively used because of the lacking support of some major companies. From 2011 onwards, the International Press Telecommunication Council (IPTC) governed the maintenance of ACAP, whose subsequent version ACAP 2.0 was rebranded to \textit{RightsML}.\footnote{\url{https://iptc.org/std/RightsML/2.0/RightsML_2.0-specification.html}} This protocol was tailored to the domain of digital news media and not commonly used in the broad context of web crawling~\cite{Initiatives}.


The ambition of the former efforts by the publishing industry partially overlaps with the scope of C2PA (Coalition for Content Provenance and Authenticity).\footnote{\url{https://c2pa.org}} This recent initiative establishes new technical standards for content authentication that verify the origin and history of web data. Among others, the C2PA manifesto specifies so-called \textit{Training And Data Mining Assertions}. These assertions can be associated to individual web assets and communicate the right to use the asset for Data Mining and the training of AI models.

\subsection{Relevant metadata standards}
In the last two decades, the domain of Semantic Web has yielded numerous metadata standards for the machine-readable annotation of web documents. These efforts have resulted in several formats for expressing copyright and licensing information. Most notably, markups like the Dublin Core Metadata Initiative (DCMI) Terms,\footnote{\url{https://www.dublincore.org/specifications/dublin-core/dcmi-terms/}} the WHATWG-specified Meta Extension \texttt{rights-standard},\footnote{\url{https://wiki.whatwg.org/wiki/MetaExtensions}} the \texttt{copyright} Meta Tag~\footnote{\url{https://www.metatags.org/all-meta-tags-overview/meta-name-copyright/}} or the \texttt{license} Microformat~\footnote{\url{https://microformats.org/wiki/rel-license}} are used to specify relevant meta information as structured data within the web page. Moreover, the WHATWG-specified \texttt{work}\footnote{\url{https://html.spec.whatwg.org/multipage/microdata.html\#licensing-works}} Microdata type and the \textit{Creative Commons Rights Expression Language (ccREL)}~\footnote{\url{https://opensource.creativecommons.org/ccrel/}} allow authors to include license information directly in the body of HTML documents.

A major drawback of many of the numerous metadata standards is the lacking adoption among wider parts of the web community. Such an ill fate was also bestowed on \textit{Do Not Track (DNT)}, whose community group publicly announced its termination in 2019. DNT and its spiritiual successor \textit{Global Privacy Control (GPC)}~\footnote{\url{https://globalprivacycontrol.org}} are designed as optional HTTP request headers that allow internet users to opt out from website tracking. Similar to ACAP, the DNT initiative was promising, but eventually abandoned by the major players. This leaded to a lack of incentives for websites to respect it, resulting in DNT's inevitable failure.

\section{Recent ad hoc standards}

Following the recent advancements in generative AI technology, there has been an increased focus on providing opt-out mechanisms for content creators. These mechanisms, conceptually similar to the \textit{Do Not Track} header, may act as a safeguard for web users.

The challenge now lies in establishing a commonly accepted technical format for the opt-out. Various protocols and metadata standards, including ACAP, RightsML, C2PA, \texttt{rights-standard}, \texttt{copyright}, and \texttt{license}, present viable but overelaborate solutions to this issue. The past few months have however yielded new, more practical techniques specifically designed to address opt-outs from the training and inference of generative AI models. This section will introduce and evaluate these ad hoc standards based on a structured schema. Moreover, an empirical study will assess the current web adoption rate of these appoaches.

\begin{figure*}[t]
\centering
\includegraphics[width=0.95\linewidth]{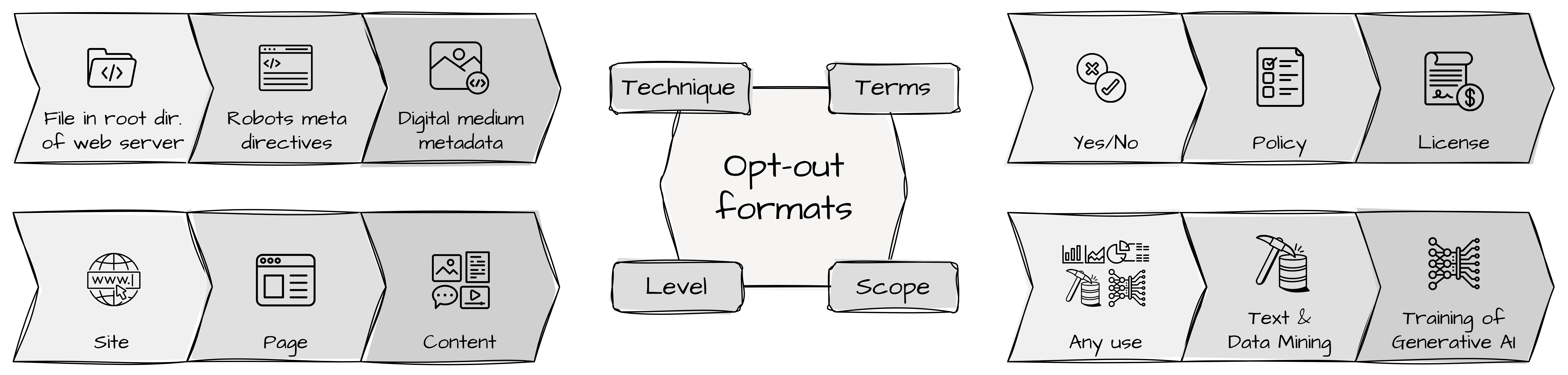}
\caption{Evaluation schema of opt-out formats}
\label{fig:opt-out-format-schema}
\end{figure*}

\subsection{Evaluation schema}
The structured schema illustrated in Figure~\ref{fig:opt-out-format-schema} guides our evaluation of technical ad hoc standards in the context of web publishers control. It comprises the following four criteria:
\vspace{-0.1cm}
\begin{itemize}
\item \textit{Technique} defines the technical means through which a rightsholder's preferences are communicated. The three distinct techniques each cater to different scenarios and technical skills a publisher may have.
\item \textit{Level} indicates the degree of granularity at which the rightsholder's preferences are specified, ranging from Site-, Page-, to Content-level. This differentiation introduces a clear hierarchy in opt-out directives: a Content-level specification is prioritized over those made at the Page- or Site-level. Additionally, the ability of a rightsholder to modify settings across these levels may be limited; for example, on a collaborative or social media platform, a rightsholder might not be allowed to alter the entire site configuration, but rather more granular elements like paragraphs, images, or individual web pages.
\item \textit{Terms} elaborates on the terms and conditions associated with the opt-out. This can be a simple binary choice (either permitting or prohibiting the use of data) or a more detailed license agreement specifying conditions of data use, such as limitations to non-commercial purposes, financial compensation, etc. While policies themselves do not impose obligations or conditions, they facilitate the automated acquisition of online licenses. The W3C-standardized Open Digital Rights Language (ODRL) is the most prominent format purposed for automated license acquisition.
\item The \textit{Scope} of the opt-out ranges from a complete ban on any form of data utilization to specific use cases, like Text \& Data Mining or the development of AI tools. For instance, site owners may wish to restrict the use of their content in AI/ML applications while still enabling web indexing and search engine discoverability.
\end{itemize}

\subsection{Overview}
The following section presents six recent proposals that allow publishers to opt out from ML training.

\subsubsection{robots.txt}

\begin{tabular}{lp{12cm}}
\textbf{Technique:} & TXT file in the web server's root directory \\
\textbf{Level:} & Site-level \\
\textbf{Terms:} & Binary signal (\texttt{allow}\,/\,\texttt{disallow}) \\
\textbf{Scope:} & Any use
\end{tabular}
\vspace{0.25cm}

The initial approach enhances the well-established Robots Exclusion Protocol by strictly adopting its basic \texttt{allow} and \texttt{disallow} commands, as detailed in the IETF RFC document for REP~\cite{Koster2022}. Below is an example of a robots.txt file section, purposefully designed to prevent the crawling of any image media content:
\vspace{-0.1cm}
\begin{verbatim}
    user-agent: *
    disallow: *.bmp           disallow: *.png           disallow: *.jpeg
    disallow: *.gif           disallow: *.svg           disallow: *.webp
    disallow: *.ico           disallow: *.tif           disallow: *.tiff
\end{verbatim}

This snippet was generated by an online tool of the initiative \textit{Spawning}, which advocates against unauthorized AI Data Mining.\footnote{\url{https://site.spawning.ai/spawning-ai-txt}} Spawning provides a web service featuring a simple generator for \texttt{disallow} directives. Its aim is to protect specific media content types from being crawled. It thus offers website administrators a basic mechanism to block the exploitation of their site's texts, images, or videos. However, instructions in the robots.txt file do not allow for the definition of specific scopes, revealing a significant limitation of the current REP standard. Moreover, implementing more complex restrictions leads to an increase in the file's size, making it overly verbose for scenarios such as excluding only certain text or video files. As a result, the robots.txt file may become too lengthy and intricate for administrators to manage it effectively, contradicting the REP guideline, which suggests keeping the file size under 500 KiB.

\subsubsection{Usage-specific agent names}

\begin{tabular}{lp{12cm}}
\textbf{Technique:} & \textcolor{gray}{$\rightarrow$ robots.txt} \\
\textbf{Level:} & \textcolor{gray}{$\rightarrow$ robots.txt} \\
\textbf{Terms:} & \textcolor{gray}{$\rightarrow$ robots.txt} \\
\textbf{Scope:} & Standard user agent: Web Indexing \\
 & Extended user agent: Training of ML models
\end{tabular}
\vspace{0.25cm}

As previously discussed, the Robots Exclusion Protocol does not differentiate between scenarios of data use. To address this limitation, some web crawl operators have introduced specific product tokens for particular data use cases. Notably, Google has unveiled a new user agent named \texttt{Google-Extended},\footnote{\url{https://blog.google/technology/ai/an-update-on-web-publisher-controls/}} which joins the lineup of 14 identifiers for Google's crawling mechanisms, including \texttt{Googlebot}, \texttt{AdsBot-Google}, \texttt{FeedFetcher-Google}, among others. When \texttt{Google-Extended} is specified to restrict access to certain website sections, Google ensures that the blocked content will not contribute to the enhancement of its flagship AI products, Bard and Vertex AI.

Such usage-specific product tokens provide a means to opt out from particular data applications, such as the development of certain AI tools, without completely prohibiting \textit{all} crawling activities. However, as Google pointed out in their statement, it is necessary ``to explore additional machine-readable approaches to choice and control for web publishers''. The company's VP of Trust express the legitimate concern that ``as AI applications expand, web publishers will face the increasing complexity of managing different uses at scale''. Although Google's market influence probably persuades stakeholders to adopt their extended user agent, this approach does not yield a sufficient solution in the long term. The necessity to declare multiple user agents, each representing a different data use by individual crawl operators, possibly results in an excessively complicated set of directives in the robots.txt file, placing an undue burden on webmasters.

\subsubsection{learners.txt}

\begin{tabular}{lp{12cm}}
\textbf{Technique:} & Two TXT files in the web server's root directory \\
\textbf{Level:} & \textcolor{gray}{$\rightarrow$ robots.txt} \\
\textbf{Terms:} & \textcolor{gray}{$\rightarrow$ robots.txt} \\
\textbf{Scope:} & robots.txt: Web Indexing \\
 & learners.txt: Training of ML models
\end{tabular}
\vspace{0.25cm}

The subsequent suggestion, introduced in July 2023, does not modify but rather replicates the robots.txt file~\cite{Ippolito2023}. It introduces a second file, similar to robots.txt, named \texttt{learners.txt} on the web server. The instruction set within both files remains the same (\texttt{allow}\,/\,\texttt{disallow}), intentionally avoiding increased complexity for webmasters and preventing any confusion. It allows webmasters to distinctly separate instructions intended for general web search crawling from those targeting AI/ML data collection, dividing them into two separate documents. For example, a traditional search engine crawler would adhere to the Robots Exclusion Protocol, while a tool designed for compiling ML training data would follow the directives specified in the `learners' file. A significant downside of this proposal is the necessity for an additional retrieval action by the crawler. Given the learners.txt file's potential low adoption rate initially, it is likely that its directives may not be widely recognized, rendering the learners' instructions largely unnoticed.

\subsubsection{NoAI, NoArchive and NoCache Meta Tags}

\begin{tabular}{lp{12cm}}
\textbf{Technique:} & Robots meta directives \\
\textbf{Level:} & Page-level \\
\textbf{Terms:} & Binary signal (e.g. \textit{no tag}\,/\,\texttt{noai}) \\
\textbf{Scope:} & Training of ML models
\end{tabular}
\vspace{0.25cm}

While not officially part of the REP specification, the robots HTML meta tags and X-Robots-Tag HTTP response header are widely utilized methods for conveying machine-readable instructions to web crawlers. In November 2022, the US-based online art community \textit{DeviantArt} introduced the directives \texttt{noai} and \texttt{noimageai} for the exclusion of AI training.\footnote{\url{https://www.deviantart.com/team/journal/update-all-deviations-are-opted-out-of-ai-datasets-934500371}} Similarly, the search engine \textit{Mojeek} supported the idea of a \texttt{noml} meta tag,\footnote{\url{https://noml.info}} aligning with the core ideas of \texttt{noai} and \texttt{noimageai}.

The launch of \texttt{noai} marked an early attempt to enhance control for web publishers over their content's use. However, its impact within the Web Search and Publishing sectors has largely been symbolic, with limited practical adoption beyond the initially involved platforms. This limited usage is partly due to the lack of support from key industry players, who have proposed their own mechanisms instead. For example, Microsoft Bing has announced to utilize the existing meta tags \texttt{noarchive} and \texttt{nocache} in their approach to offering webmasters more control over their content.\footnote{\url{https://blogs.bing.com/webmaster/september-2023/Announcing-new-options-for-webmasters-to-control-u}} According to Bing, using the \texttt{nocache} tag ensures that the content is excluded from training Microsoft’s generative AI models, and the \texttt{noarchive} tag prevents content from being referenced in Bing Chat responses. The reinterpretation of the existent instructions, however, presents a significant challenge for webmasters due to the ambiguity of the two tags and the increased complexity of managing a multitude of tags.

\subsubsection{NO\_TRAIN Metadata Field}

\begin{tabular}{lp{12cm}}
\textbf{Technique:} & Digital image metadata \\
\textbf{Level:} & Content-level \\
\textbf{Terms:} & Binary signal (\texttt{DO\_TRAIN}\,/\,\texttt{NO\_TRAIN}) \\
\textbf{Scope:} & Training of ML models
\end{tabular}
\vspace{0.25cm}

A further strategy for enhancing control over the use of web resources involves embedding meta information directly within the resources. The \textit{DoNotTrain} Metadata Standard introduces a concept in this vein, featuring a \texttt{NO\_TRAIN} tag specifically for image metadata~\cite{Ippolito2023}. This tag is designed to clearly indicate that the use of the image (or any digital medium it is applied to) for Machine Learning training is explicitly prohibited. Importantly, the \texttt{NO\_TRAIN} metadata can be incorporated into various file types prevalent on the web, including audio, video, and text files, although it is most directly applicable to images. Given that the first accusations on potential copyright violations of AI companies are related to images~\cite{Samuelson2023}, starting with image metadata for these efforts is a logical choice, likely to gain significant attention and support from the practitioners community.

\subsubsection{TDM Reservation Protocol}

\begin{tabular}{lp{12cm}}
\textbf{Technique:} & JSON file in the web server's \texttt{.well-known} directory; Robots meta directives \\
\textbf{Level:} & Site- or Page-level \\
\textbf{Terms:} & Binary signal (no tag\,/\,tdm-reservation); TDM Policy \\
\textbf{Scope:} & Text \& Data Mining
\end{tabular}
\vspace{0.25cm}

The \textit{TDM Reservation Protocol (TDM Rep)}, a web standard endorsed by the W3C, detaches from the Robots Exclusion Protocol to be more fit to today's increased awareness on data sovereignty~\cite{TDMRep}. This protocol grants publishers the ability to specify their preferences regarding the Text \& Data Mining of online resources under their control. Finalized in 2022, it aligns with the regulations of the EU's DSM Directive of 2019, which recommends that rightsholders may opt out of TDM activities by appropriately asserting their rights in a machine-readable format. Recognizing the lack of such formats for TDM, the creators of this protocol introduced a new standard that meets the needs of web publishers while remaining straightforward and practical.

TDM Rep allows webmasters to declare their TDM rights preferences by marking individual documents with the \texttt{tdm-reservation} tag. This tag is binary: a value of 1 indicates a reservation of rights, while 0 signifies no reservation, permitting web agents to mine the content without further consultation with the rightsholder, as default under Article 4 of the DSM Directive. Additionally, the protocol outlines the use of the \texttt{tdm-policy} tag for directing to a TDM Policy. This tag is linked to a URL that hosts an ODRL policy document in JSON format, facilitating the automatic licensing of web resources by detailing the rightsholder's contact information and the terms and conditions of use.

The protocol outlines three methods for communicating the decisions of rights holders. The first two mirror the use of robots HTML meta tags and X-Robots-Tag HTTP response headers. The third involves creating a JSONL document named \texttt{tdmrep.json}, placed within the \texttt{.well-known} directory of the web server. This document lists JSON objects, each representing a rule with properties for \texttt{location}, \texttt{tdm-reservation}, and optionally, \texttt{tdm-policy}. The \texttt{location} property, similar to the directives in a robots.txt file, specifies a URL path within the site.

\subsection{Empirical study}
This study assesses the adoption rate of the previously discussed ad hoc standards. Overall, we have analyzed 60 million regular web pages along with their respective tdmrep.json files, if existent, as well as 42 million robots.txt files. The documents were sourced from the web archive of Common Crawl,\footnote{\url{https://commoncrawl.org}} offering a broad and randomly selected cross-section of the internet. These files were all collected in November and December 2023. Further details on the experimental framework and comprehensive crawl statistics can be found in~\cite{Dinzinger2024}.

\paragraph{REP}
Around 56.0 \% of crawled websites currently provide a valid robots.txt file. This figure has remained relatively unchanged for the past eight years, reflecting the Robots Exclusion Protocol's solidified role as the primary regulation tool for web crawlers. The \texttt{robots} HTML meta tag is present in about 52.7~\% of web pages, while the X-Robots-Tag header appears in merely 0.6~\% of the HTTP responses we analyzed.

\paragraph{User agents}
Specific user agents that are used to exclude AI training, such as Google-Extended, GPTBot or CCBot, have seen considerable adoption across websites. Google-Extended, for instance, is mentioned in over 650,000 robots.txt files of the 42 million we examined; a remarkable increase since its launch in September 2023. Notably, Google-Extended frequently appears alongside a \textit{disallow all} directive (\texttt{Disallow:\,/}). As of December 2023, approximately 653,800 websites categorically block the web agent, accounting for 99.9~\% of the around 654,300 robots.txt files that mention Google-Extended and 1.6~\% of all 42 million robots.txt files analyzed. Most prominently, news websites like lemonde.fr, washingtonpost.com and nytimes.com, which are currently in a lawsuit against OpenAI, use these user agents to opt out from AI training. Search Engine bots such as Googlebot and Bingbot, on the other hand, are generally more welcomed by webmasters~\cite{Dinzinger2024}. This trend suggests a preference among many site owners to be indexed by major search engines while excluding their content from AI model training and inference.

\paragraph{Meta tags}
The introduction of \texttt{noai}, \texttt{noimageai}, and \texttt{noml} meta tags has yet to gain significant traction within the online community. These tags were detected on the HTML pages of only 82 out of around 1.4 million distinct hosts. The \texttt{noarchive} and \texttt{nocache} tags, however, are utilized by up to 1.27~\% of web pages, leading to an exclusion from Bing Chat's response generation and training processes. Microsoft’s decision thus remains questionable as the company profits from the low adoption of these meta tags among websites because consequently more data is available for them to run and improve their AI products. For that cause, they accept the increased ambiguity that comes along with reinterpreting the existing robots instructions for this new purpose.

\paragraph{TDM Rep}
The adoption of the TDM Reservation Protocol is still low. By January 2024, we found a minuscule number of hosts (45) serving a tdmrep.json file within the \texttt{.well-known} directory of their servers. Around 60 domains have implemented TDM Rep by including the \texttt{tdm-reservation} tag. Particularly French websites, e.g. lefigaro.fr, appear to be leading the charge in committing to this new protocol.

\section{Conclusion}

This survey examines the current discussion on web publishers control, which demands new web standards for the opt-out from AI training. In this course, we want to highlight both legal frameworks and existing technical standards. Regarding the legal aspects, the protection of intellectual property is well-defined in the EU and US law. These regulations generally apply to web crawling and scraping, requiring in any case a justification against the copyright law. Crawling for regular web indexing is mostly considered legitimate because it is beneficial to all parties involved. Hence, the overall assessment poses an implied consent by the site owners, profiting from having their website indexed and included in search results. In 2019, the EU enacted the DSMD that regulates Text \& Data Mining and therefore also any web search-related crawling, rendering the common practice of implied consent redundant. According to Art. 4 DSMD, reproductions of web content are by default permissible but rightsholders are given the opportunity to reserve the use of their works using a machine-readable opt-out. Additionally, TDM for the purposes of scientific research is exempted from the opt-out regulation and thus always permissible.

The recent advancements on generative AI have yielded numerous ad hoc standards, accommodating web publishers' wish for more data sovereignty. These techniques are specifically designed to address the opt-out from training and inference of generative AI models. Our paper evaluates the in this regard most relevant proposals based on a structured schema. The proposals build off of conventional protocols, such as the Robots Exclusion Protocol, or bring forward new initiatives, such as the TDM Reservation Protocol. They were introduced by the online art, publishers and researchers community as well as by Search Engines like Google, Microsoft Bing and Mojeek. Most of the ad hoc standards have - at least at the beginning - only a small impact beyond their symbolic radiance. Only the usage-specific web agents like Google-Extended or GPTBot, are relatively frequent in robots.txt files, expressing the publishers' intention of excluding AI training through the Robots Exclusion Protocol.

In summary, the presented approaches are either idealistic and thus poorly adopted, such as the \texttt{noai}/\texttt{noml} meta tag or the learners.txt file, or they are specific to certain AI applications, such as the Google-Extended user agent for Google's Bard and Vertex AI, the user agents GPTBot and ChatGPT-User for OpenAI's products, and the \texttt{nocache} and \texttt{noarchive} meta tags for Microsoft's Bing Chat. Furthermore, it will always remain the possibility for AI companies to not crawl themselves, but use existing open data such as Common Crawl, effectively evading any site-level control measures based on robots.txt. As a result, webmasters are faced with an overwhelming task to implement \textit{all} these technical micro-standards or at least consider the most crucial of them, in order to protect their data from any unwanted AI/ML-related use.

\begin{acknowledgments}
This work has received funding from the European Union's Horizon Europe research and innovation program under grant agreement No 101070014 (OpenWebSearch.EU, \url{https://doi.org/10.3030/101070014}). 
\end{acknowledgments}

\bibliography{sources}


\end{document}